# PhishNET: A Phishing Websites Detection Tool

## A PROJECT REPORT

Submitted in partial fulfillment of the requirement for the award of the degree

of

### BACHELOR OF TECHNOLOGY

in

### COMPUTER SCIENCE AND ENGINEERING

**SUBMITTED BY**

Arshpreet Singh Sohal (20103030)

Deepakmoney Banga (20103047)

Kevin Antony (20103078)

Under the supervision of

**Dr. Prashant Kumar**
**Assistant Professor**

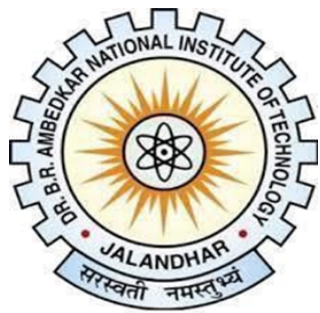

**Department of Computer Science and Engineering**
**Dr. B. R. Ambedkar National Institute of Technology Jalandhar**
**-144008, Punjab (India)**
**May 2024**

# CANDIDATES' DECLARATION

We hereby certify that the work presented in this project report entitled "**PhishNET: A Phishing Websites Detection Tool**" in partial fulfillment of the requirement for the award of a Bachelor of Technology degree in Computer Science and Engineering, submitted to the Dr. B R Ambedkar National Institute of Technology, Jalandhar is an authentic record of our own work carried out during the period from July 2023 to May 2024 under the supervision of Dr. Prashant Kumar, Assistant Professor, Department of Computer Science & Engineering, Dr. B R Ambedkar National Institute of Technology, Jalandhar.

We have not submitted the matter presented in this report to any other university or institute for the award of any degree or any other purpose.

Date: 29th May, 2024

Submitted by
Arshpreet Singh Sohal (20103030)
Deepakmoney Banga (20103047)
Kevin Antony (20103078)

This is to certify that the statements submitted by the above candidates are accurate and correct to the best of our knowledge and are further recommended for external evaluation.

Dr. Prashant Kumar, Supervisor                 Dr. Rajneesh Rani
Assistant Professor                                      Head and Associate Professor
Deptt. of CSE                                                 Deptt. of CSE

# **ACKNOWLEDGEMENT**

It is true that hundreds of people work behind the scenes for the success of a play. The end result of the PhishNet project required a lot of guidance and help from many people, and our group was very fortunate to receive this support during the course of the project. Whatever we have achieved today is only due to such supervision and assistance, and we thank them from the bottom of our hearts.

We would like to express our deepest gratitude to our project mentor, Dr. Prashant Kumar, Assistant Professor, who believed in our ideas and suggested new approaches when needed. He fully supported us in solving our problems.

We would also like to express our deepest gratitude to Dr. Rajneesh Rani, Head of the Department of Computer Science and Engineering, for her direct and indirect support.

We are grateful to Dr. Aruna Malik, Coordinator Major Project, for providing us with mentors and all other support.

We are extremely thankful to have constant encouragement and guidance from all the faculty members of the Department of Computer Science & Engineering. We would also like to express our sincere thanks to all laboratory staff for their timely support.

Thank you.

[Arshpreet, Deepak, Kevin]

# **ABSTRACT**

PhisNet is an innovative web-based application designed to detect phishing websites through the application of advanced machine learning technologies. This project addresses the challenges faced by individuals and organizations in identifying and preventing phishing attacks. Built on a robust artificial intelligence framework, PhisNet employs various machine learning algorithms and feature extraction techniques using Python to ensure high accuracy and efficiency in phishing detection.

The project begins with the collection and preprocessing of a comprehensive dataset of URLs, including both phishing and legitimate websites. Significant features are extracted from these URLs, such as URL length, presence of special characters, and domain age, to train the model effectively. Multiple machine learning algorithms, including logistic regression, decision trees, and neural networks, are evaluated for their performance in detecting phishing websites. The model is meticulously trained to optimize performance metrics such as accuracy, precision, recall, and the F1 score, ensuring reliable detection of both common and sophisticated phishing tactics.

PhisNet's full stack web application is developed using React.js, a powerful frontend framework that allows for client-side rendering and seamless integration with backend services. This choice facilitates the creation of a responsive and user-friendly interface. Users can input URLs and receive immediate predictions with confidence scores, supported by a robust backend infrastructure that processes data and provides real-time results. The model is deployed using Google Colab and AWS EC2, chosen for their computational power and scalability, ensuring the application remains accessible and functional under varying loads of user requests.

In conclusion, PhisNet represents a significant technological advancement in cybersecurity, demonstrating the effective application of machine learning and web development technologies to enhance user security. It empowers users to take preventive measures against phishing attacks and highlights the potential of AI in transforming cybersecurity.

# PLAGIARISM REPORT

We have checked plagiarism for our Project Report for our project a **Turnitin.** We are thankful to our mentor Dr. Prashant Kumar for guiding us at this. Below is the digital receipt. The Plagiarism is approximately 10%.

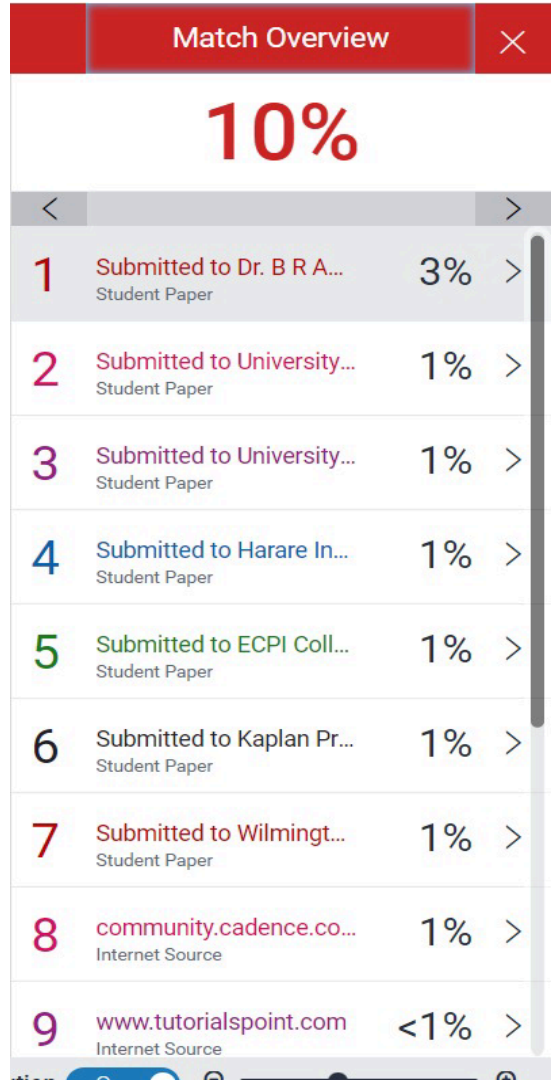

# LIST OF FIGURES

# LIST OF ABBREVIATIONS

SVM: Support Vector Machine

RF: Random Forest

NN: Neural Network

kNN: k-Nearest Neighbors

AWS EC2: Amazon Web Services Elastic Compute Cloud

URL: Uniform Resource Locator

IP: Internet Protocol

HTTP: Hypertext Transfer Protocol

HTTPS: Hypertext Transfer Protocol Secure

CSV: Comma-Separated Values

DNS: Domain Name System

AWS: Amazon Web Services

# **TABLE OF CONTENTS**

# CHAPTER 1
# INTRODUCTION

## 1.1 Background

In the realm of cybersecurity, phishing poses a significant threat, with attackers continuously evolving tactics to deceive users into divulging sensitive information. Phishing attacks, disguised as legitimate entities, aim to steal personal data such as usernames, passwords, and financial information [7]. Despite awareness efforts, individuals and organizations remain vulnerable to these sophisticated attacks.

PhisNet emerges as a solution to combat these challenges, offering an online platform powered by advanced machine learning technologies. By extracting features from URLs and leveraging machine learning models, PhisNet classified websites as either phishing or legitimate, providing users with a tool to identify potential threats proactively.

## 1.2. Literature Survey

We have conducted a brief literature survey to contextualize our research within the domain of advanced machine learning algorithms and feature extraction techniques. This survey aims to identify key methodologies and trends in existing studies, providing insights that inform our approach. By reviewing relevant literature, we aim to bridge current knowledge gaps and contribute to the enhancement of user security.

1. **Detecting Phishing Websites Using Machine Learning** [1][6]**:**
   This paper presents an intelligent system implemented as a browser extension for detecting phishing websites, employing supervised learning with the Random Forest technique. The system automatically alerts users when encountering potential phishing sites, enhancing internet browsing security.
2. **Phishing Website Detection From Url Using Machine Learning** [2][4]**:**

This research aims to enhance defense mechanisms against phishing by exploring diverse approaches for website categorization. The system uses machine learning techniques, including decision tree, support vector machine (SVM), Naïve Bayesian classifier, and neural network, to detect phishing websites based on their URLs.

3. **Phishing Website Detection Using Different Machine Learning Algorithms [3][5]:**

This paper aims to present an application to detect phishing websites from their urls using a stacking model. It uses two features, encompassing both the strongest and weakest attributes, and is proposed and subjected to principal component analysis(PCA). Diverse machine learning algorithms, such as random forest (RF), neural network (NN) etc are used.

## 1.3. Problem Statement and its Necessity

PhisNet addresses several critical issues:

1. **Unavailability of Advanced Security Solutions**:

Many entities lack access to sophisticated tools for detecting phishing websites, leaving them vulnerable to cyberattacks. PhisNet offers an accessible and effective solution to identify phishing threats.

2. **Early Detection:**

Proactive identification of phishing attempts is crucial to mitigate risks. PhisNet facilitates early detection, reducing the likelihood of data breaches and financial losses.

3. **Supporting IT Professionals:**

IT professionals, especially in smaller organizations, may lack resources to implement robust security measures. PhisNet provides a reliable tool to aid in phishing detection, bolstering overall cybersecurity posture.

### 1.4. Motivation

- Access to effective solutions for detecting phishing websites is often limited and costly, requiring specialized expertise. PhisNet was developed to democratize phishing detection, making it accessible and user-friendly for individuals and organizations alike.

- By leveraging machine learning algorithms to analyze URL features, PhisNet empowers users to assess the likelihood of a URL being a phishing site. This innovative approach offers an affordable and practical means to enhance cybersecurity, even for users without specialized training or resources.

### 1.5. Feasibility : Non-Technical and Technical

Assessing the feasibility of the project from various standpoints:

**TECHNICAL**:
The project leverages powerful programming languages such as Python, along with comprehensive support for machine learning algorithms and cloud resources like Google Colaboratory and AWS EC2. React.js, a frontend framework, facilitates the development of responsive web applications.

**SOCIAL**:
Currently, there is no widely adopted application addressing phishing detection using advanced machine learning techniques.

**ECONOMICAL**:
The project's development expenses are minimal, utilizing open-source libraries and publicly available datasets for model training.

**SCOPE**:

PhisNet aims to assist users, including IT professionals and individuals, by providing preliminary assessments of potentially malicious URLs.

## 1.6 Research Objectives

PhisNet aims to revolutionize phishing website detection by leveraging advanced machine learning technologies, focusing on enhancing accessibility for individuals and organizations. Through feature extraction and machine learning algorithms, PhisNet ensures high accuracy and reliability in detecting phishing attempts, thereby transforming cybersecurity with AI-powered threat detection [8].

# CHAPTER 2
# PROPOSED SOLUTION

PhisNet is a comprehensive web application and Chrome extension designed to combat phishing threats through advanced machine learning algorithms. By leveraging these algorithms, PhisNet empowers users to identify and mitigate potential phishing attacks in a quick and efficient manner, thereby enhancing cybersecurity.

**2.1 PhisNet Web Application**

The PhisNet web application serves as a central hub for users to assess the legitimacy of URLs and detect phishing attempts. Key features of the web application include:

URL Analysis: Users can input URLs into the application for analysis using deep learning classification algorithms, providing a preliminary diagnosis of the likelihood of a URL being a phishing site [9].

Classification Accuracy: PhisNet uses machine learning models trained on extensive datasets to ensure high accuracy [10] in identifying phishing URLs. Users receive confidence scores for each classification, aiding in decision-making.

User-Friendly Interface: The web application gives a user-friendly interface, allowing users to easily input URLs and get the results of the analysis.

**2.2 PhisNet Chrome Extension**

In addition to the web application, PhisNet offers a Chrome extension to provide users with real-time phishing detection capabilities directly within their browser. The Chrome extension enhances user protection against phishing threats by:

**Real-Time Analysis:** The extension provides real-time analysis of URLs as users navigate the web, alerting them to potential phishing sites before they interact with them.

**Browser Integration:** PhisNet integrates seamlessly into the user's browsing experience, offering convenient access to phishing detection tools without requiring them to leave the webpage they are visiting.

**Customizable Settings:** Users can customize the extension's settings to adjust the level of sensitivity to phishing threats, tailoring the protection to their specific needs and preferences.

**2.3 Development Process**

The development of PhisNet involves several key stages:

**Data-Collection:** Gathering a comprehensive dataset of URLs, including both phishing and benign websites, to train the machine learning models.

**Model Training:** Training deep learning classification algorithms on the collected dataset to develop accurate models for identifying phishing URLs.

**Web Application Development:** Building the PhisNet web application with a focus on user experience and functionality, incorporating the trained machine learning models for URL analysis.

**Chrome Extension Development:** Designing and implementing the PhisNet Chrome extension to seamlessly integrate with users' browsing experience and provide real-time phishing detection capabilities.

**Testing and Deployment:** Thorough testing of both the web application and Chrome extension to ensure reliability, accuracy, and compatibility across different platforms and browsers, followed by deployment to production environments.

PhisNet offers several benefits to users:

**Enhanced Security:** Users can identify and avoid phishing threats, reducing the risk of becoming a victim to cyberattacks and safeguarding their personal information.

**Convenience:** The web application and Chrome extension provide convenient and accessible tools for phishing detection, empowering users to protect themselves while browsing the internet.

**Customization:** Users can customize their protection settings to align with their browsing habits and security preferences, ensuring a tailored and effective defense against phishing threats.

**Future enhancements for PhisNet may include:**

**Enhanced Machine Learning Models:** Continuously improving and fine-tuning the machine learning models to adapt to evolving phishing tactics and enhance classification accuracy.

**Expanded Browser Support:** Extending support for additional web browsers beyond Chrome to reach a broader user base and provide comprehensive protection across different platforms.

**Integration with Security Suites:** Integrating PhisNet with existing security suites and tools to offer comprehensive protection against a range of cyber threats, including phishing attacks on web.

**The following was the procedure followed to do the project:**

**Phase 1 - Collecting the Data:**

For this project, we need a bunch of urls of type legitimate (0) and phishing (1).

The collection of phishing urls is done easily because of the open source service called PhishTank. This service gives a set of legitimate and phishing URLs in csv format. For downloading the data the url is: https://www.phishtank.com/developer_info.php

For the valid URLs, We located a source that has a collection of benign, spam, phishing, malware & defacement URLs. The source of the dataset is University of New Brunswick, https://www.unb.ca/cic/datasets/url-2016.html. The number of legitimate URLs in this collection is 35,300. The URL collection is downloaded & from that, *'Benign_list_big_final.csv'* is the file of our interest. This file is thereafter uploaded to the Colab for the feature extraction.

**Phishing URLs:**

The phishing URLs are collected from the PhishTank from the link provided. The csv file of phishing URLs is obtained by using wget command. After downloading the dataset, it is loaded into a DataFrame.

|   | phish_id | url | phish_detail_url | submission_time | verified | verification_time | online | target |
|---|---|---|---|---|---|---|---|---|
| 0 | 6557033 | http://u1047531.cp.regruhosting.ru/acces-inges... | http://www.phishtank.com/phish_detail.php?phis... | 2020-05-09T22:01:43+00:00 | yes | 2020-05-09T22:03:07+00:00 | yes | Other |
| 1 | 6557032 | http://hoysalacreations.com/wp-content/plugins... | http://www.phishtank.com/phish_detail.php?phis... | 2020-05-09T22:01:37+00:00 | yes | 2020-05-09T22:03:07+00:00 | yes | Other |
| 2 | 6557011 | http://www.accsystemprblemhelp.site/checkpoint... | http://www.phishtank.com/phish_detail.php?phis... | 2020-05-09T21:54:31+00:00 | yes | 2020-05-09T21:55:38+00:00 | yes | Facebook |
| 3 | 6557010 | http://www.accsystemprblemhelp.site/login_atte... | http://www.phishtank.com/phish_detail.php?phis... | 2020-05-09T21:53:48+00:00 | yes | 2020-05-09T21:54:34+00:00 | yes | Facebook |
| 4 | 6557009 | https://firebasestorage.googleapis.com/v0/b/so... | http://www.phishtank.com/phish_detail.php?phis... | 2020-05-09T21:49:27+00:00 | yes | 2020-05-09T21:51:24+00:00 | yes | Microsoft |

Figure 2.1: Phishing URLs 1.0

There are thousands of phishing URLs in the data. The issue here is that the data is refreshed every hour. Without getting into the risk of data imbalance, I am considering a margin value of 10,000 phishing URLs & 5000 legitimate URLs.

As of now we collected 5000 phishing URLs. Now, we need to collect the legitimate URLs.

|   | phish_id | url | phish_detail_url | submission_time | verified | verification_time | online | target |
|---|---|---|---|---|---|---|---|---|
| 0 | 6485787 | https://eevee.tv/Bootstrap/assets/css/acces | http://www.phishtank.com/phish_detail.php?phis... | 2020-04-04T03:01:00+00:00 | yes | 2020-04-04T03:03:56+00:00 | yes | Other |
| 1 | 6422543 | https://appleid.apple.com-sa.pm/appleid/? | http://www.phishtank.com/phish_detail.php?phis... | 2020-02-27T17:01:01+00:00 | yes | 2020-03-17T01:50:51+00:00 | yes | Other |
| 2 | 6543602 | https://grandcup.xyz/ | http://www.phishtank.com/phish_detail.php?phis... | 2020-05-02T23:07:29+00:00 | yes | 2020-05-02T23:09:03+00:00 | yes | Steam |
| 3 | 6528783 | https://villa-azzurro.com/onedrive/ | http://www.phishtank.com/phish_detail.php?phis... | 2020-04-25T20:54:02+00:00 | yes | 2020-04-25T21:46:55+00:00 | yes | Other |
| 4 | 6498136 | http://mygpstrip.net/ii/u.php | http://www.phishtank.com/phish_detail.php?phis... | 2020-04-10T15:01:56+00:00 | yes | 2020-04-10T16:01:37+00:00 | yes | Other |

Figure 2.2: Phishing URLs 1.1

**Legitimate URLs:**

From the uploaded *Benign_list_big_final.csv* file, the URLs are loaded into a python dataframe.

As stated above, the 5000 legitimate URLs are randomly picked from the above dataframe.

|   | URLs |
|---|---|
| 0 | http://graphicriver.net/search?date=this-month... |
| 1 | http://ecnavi.jp/redirect/?url=http://www.cros... |
| 2 | https://hubpages.com/signin?explain=follow+Hub... |
| 3 | http://extratorrent.cc/torrent/4190536/AOMEI+B... |
| 4 | http://icicibank.com/Personal-Banking/offers/o... |

Figure 2.3: Legitimate URLs 1.0

**Feature Extraction:**

In this step, features for training the model are extracted from the URLs dataset.

The extracted features are categorized into

1. Address Bar based Features
2. Domain based Features

**1. Address Bar Based Features:**

Many features can be extracted that can be considered as address bar base features. Out of them, below mentioned were considered for this project.

- Domain of URL
- IP Address in URL
- Length of URL
- AtSign "@" Symbol in URL
- Depth of URL
- Redirection symbol "//" in URL
- "http/https" in Domain name
- Using URL Shortening Services used: "TinyURL"
- Prefix or Suffix "-" in Domain

**1.1. Domain of the URL**

Here, we are just extracting the domain present in the URL. This feature doesn't have much significance in the training. May even be dropped while training the model.

**1.2. IP Address in the URL**

Verifies whether the URL contains an IP address. IP addresses may appear in URLs in place of domain names. If an IP address is used as an alternative of the domain name in the URL, we can be sure that someone is trying to steal personal information with this URL.

In the event that the domain portion of the URL contains an IP address, this feature will be assigned a value of 1 (phishing) or 0 (legal).

### 1.3. "@" Symbol in URL

Checks for the presence of '@' symbol in the URL. Using "@" symbol in the URL leads the browser to ignore everything preceding the "@" symbol and the real address often follows the "@" symbol.

If the URL has '@' symbol, the value assigned to this feature is 1 (phishing) or else 0 (legitimate).

### 1.4. Length of URL

Verifies whether the '@' symbol is present in the URL. When you use the "@" symbol in a URL, the browser will ignore everything that comes before it. The actual address usually comes after the "@" symbol.

If the URL contains the '@' symbol, this feature is given a value of 1 (phishing) or 0 (legal).

### 1.5. Depth of URL

Calculates the URL's depth. This functionality uses the '/' to determine how many subpages are in the provided url.

Based on the URL, the feature's value is expressed in numbers.

### 1.6. Redirection "//" in URL

Verify that "//" appears in the URL. The user will be moved to another website if the URL path contains the slash "//." The URL's "//" location is calculated. We discover that the "//" should occur in the sixth position if the URL begins with "HTTP." On the other hand, the "//" should occur in seventh position if the URL uses "HTTPS".

When a URL contains "//" anywhere other than after the protocol, it is classified as either legitimate or phishing, with a value of 0 otherwise.

**1.7. "http/https" in Domain name**

Verifies whether "http/https" is present in the URL's domain portion. Phishers may spoof a URL by adding the "HTTPS" token to the domain portion.

The value for this feature is either 1 (phishing) or 0 (legitimate) depending on whether the URL contains "http/https" in the domain portion.

**1.8. Using URL Shortening Services "TinyURL"**

On the "World Wide Web," a technique known as "URL shortening" allows a URL to be significantly shortened while maintaining its connection to the necessary webpage. This is achieved by creating a "HTTP Redirect" on a small domain name that points to the webpage with the lengthy URL.

The value assigned to this feature is either 1 (phishing) or 0 (legal) depending on whether the URL uses shortening services.

**1.9. Prefix or Suffix "-" in Domain**

Verifying if the domain portion of the URL contains a '-'. Legitimate URLs seldom ever utilize the dash symbol. In order to give visitors the impression that they are interacting with a trustworthy website, phishers frequently append prefixes or suffixes to the domain name, separated by a (-).

A value of 1 (phishing) or 0 (legal) is assigned to this feature if the URL contains the '-' symbol in the domain portion of the URL.

**2. Domain Based Features:**

Many features can be extracted that come under this category. Out of them, below mentioned were considered for this project.

- DNS Record
- Website Traffic
- Age of Domain
- End Period of Domain

**2.1. DNS Record**

When it comes to phishing websites, either the WHOIS database does not recognize the stated identity or there are no records for the hostname. The value assigned to this characteristic is either 1 (phishing) or 0 (legal) depending on whether the DNS record is empty or cannot be located.

**2.2. Web Traffic**

This feature counts how many people visit the website and how many pages they see in order to gauge its popularity. Nevertheless, as phishing websites are transient, the Alexa database might not identify them (Alexa the Web Information Company, 1996). Upon analyzing our dataset, we discovered that, in the worst-case situation, reputable websites were placed in the top 100,000. Moreover, the domain is labeled as "Phishing" if it receives no traffic or is not identified by the Alexa database.

This feature has a value of 1 (phishing) if the domain rank is less than 100,000, and 0 (legal) otherwise.

**2.3. Age of Domain**

It is possible to extract this feature from the WHOIS database. The majority of phishing websites only exist temporarily. For the purposes of this project, a legal domain must be at least 12 months old. Here, age simply refers to the difference between creation and expiration times.

The value of this feature is 1 (phishing) if the domain is older than 12 months, and 0 (legal) otherwise.

**2.4. End Period of Domain**

It is possible to extract this feature from the WHOIS database. The remaining domain time for this feature is determined by subtracting the current time from the expiration time. For this project, the lawful domain has an end period of no more than six months.

This feature's value is 1 (phishing) if the domain's expiration period is longer than six months, and 0 (legal).

**Final Dataset**

We created two dataframes with features of authentic and phishing URLs in the section above. For the machine learning training that is being completed in a different notebook, we will now merge them into a single dataframe and export the data to a CSV file.

**6. Conclusion**

With this the objective of this notebook is achieved. We finally extracted 18 features for 10,000 URLs which have 5000 phishing & 5000 legitimate URLs.

**Phase 2 - Loading Data:**

The features are extracted and stored in the csv file. The working of this dataset can be seen in the 'Phishing Website Detection_Feature Extraction.ipynb' file.

The results csv file is uploaded to this notebook and stored in the dataframe.

Our dataset looks like this:

|   | Domain | Have_IP | Have_At | URL_Length | URL_Depth | Redirection | https_Domain | TinyURL | Prefix/Suffix | DNS_Record | Web_Traffic | Domain_Age | Domain_End | iFrame | Mouse_Over | Right_Click | Web_Forwards | Label |
|---|---|---|---|---|---|---|---|---|---|---|---|---|---|---|---|---|---|---|
| 0 | graphicriver.net | 0 | 0 | 1 | 1 | 0 | 0 | 0 | 0 | 0 | 1 | 1 | 1 | 0 | 0 | 1 | 0 | 0 |
| 1 | ecnavi.jp | 0 | 0 | 1 | 1 | 1 | 0 | 0 | 0 | 0 | 1 | 1 | 1 | 0 | 0 | 1 | 0 | 0 |
| 2 | hubpages.com | 0 | 0 | 1 | 1 | 0 | 0 | 0 | 0 | 0 | 1 | 0 | 1 | 0 | 0 | 1 | 0 | 0 |
| 3 | extratorrent.cc | 0 | 0 | 1 | 3 | 0 | 0 | 0 | 0 | 0 | 1 | 0 | 1 | 0 | 0 | 1 | 0 | 0 |
| 4 | icicibank.com | 0 | 0 | 1 | 3 | 0 | 0 | 0 | 0 | 0 | 1 | 0 | 1 | 0 | 0 | 1 | 0 | 0 |

Figure 2.4: Extracted Features

**Visualizing the dataset**

Few plots and graphs are displayed to find how the data is distributed and how features are related to each other.

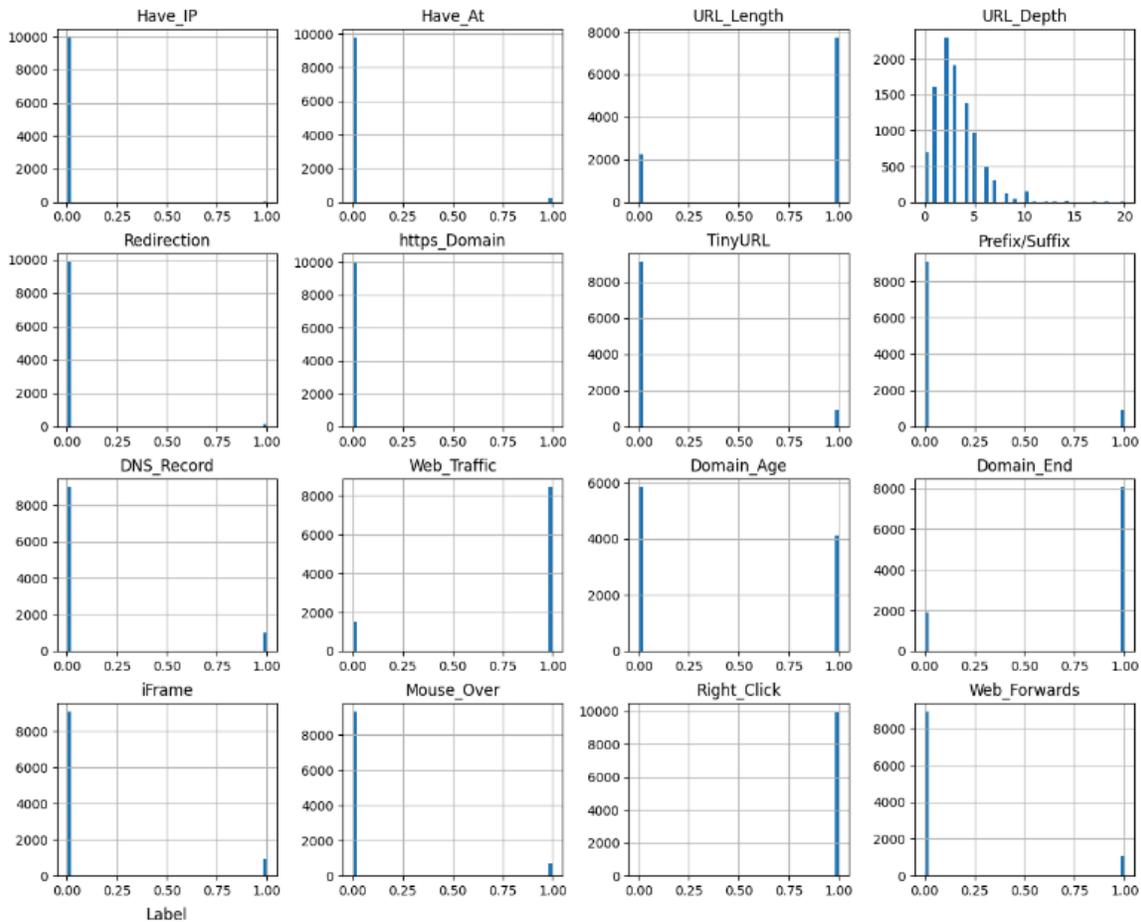

Figure 2.5: Visualization of Features

**Data Pre-processing & EDA**

Here, we clean the data by applying data preprocessing techniques and transform the data to use it in the models.

Figure 2.6: Cleaning Data

All of the data, with the exception of the "Domain" and "URL_Depth" columns, is composed of zeros and ones, as the above-mentioned result illustrates. The training of the machine learning model is unaffected by the Domain column. Eliminating the 'Domain' column from the dataset.

After that, we have 16 features and a goal column. The maximum value for 'URL_Depth' is 20. Our belief is that this column does not need to be changed.

The retrieved features of the datasets containing authentic and phishing URLs are simply concatenated without any manipulation in the feature extraction file. The top 5000 rows of authentic url data and the lowest 5000 rows of phishing url data were the outcome of this.

We must shuffle the data in order to balance the distribution after dividing it into training and testing sets. This even evades the case of overfitting while model training.

Figure 2.7: Program & Output 1.0

From the above execution, it is clear that the data doesn't have any missing values.

By this, the data is thoroughly preprocessed & is ready for training.

**Machine Learning Models & Training**

It is evident from the dataset above that this machine learning activity has to be supervised. Classification and regression are the two main categories of supervised machine learning issues.

The input URL for this data set is categorized as either authentic (0) or phishing (1), which raises classification issues. To train the dataset in this notebook, the following supervised machine learning models (classification) were taken into consideration:

- Decision Tree
- Random Forest
- Multilayer Perceptrons
- XGBoost
- Autoencoder Neural Network
- Support Vector Machines

**1. Decision Tree Classifier**

Models for regression and classification that are frequently used include decision trees. They basically pick up a hierarchy of if/else questions that lead to a choice. Understanding a decision tree entails understanding the order in which the if/else questions lead to the correct response the quickest.

These inquiries are referred to as tests in the context of machine learning (not to be confused with the test set, which is the data we use to assess the generalizability of our model). The algorithm looks over every test that might be done and selects the most useful one about the target variable in order to construct a tree.

```python
# Decision Tree model
from sklearn.tree import DecisionTreeClassifier

# instantiate the model
tree = DecisionTreeClassifier(max_depth = 5)
# fit the model
tree.fit(X_train, y_train)
```

```
▼       DecisionTreeClassifier
DecisionTreeClassifier(max_depth=5)
```

```python
#predicting the target value from the model for the samples
y_test_tree = tree.predict(X_test)
y_train_tree = tree.predict(X_train)
```

Figure 2.8: Program & Output 1.1

```python
#computing the accuracy of the model performance
acc_train_tree = accuracy_score(y_train,y_train_tree)
acc_test_tree = accuracy_score(y_test,y_test_tree)

print("Decision Tree: Accuracy on training Data: {:.3f}".format(acc_train_tree))
print("Decision Tree: Accuracy on test Data: {:.3f}".format(acc_test_tree))
```

```
Decision Tree: Accuracy on training Data: 0.813
Decision Tree: Accuracy on test Data: 0.814
```

```python
#checking the feature improtance in the model
plt.figure(figsize=(9,7))
n_features = X_train.shape[1]
plt.barh(range(n_features), tree.feature_importances_, align='center')
plt.yticks(np.arange(n_features), X_train.columns)
plt.xlabel("Feature importance")
plt.ylabel("Feature")
plt.show()
```

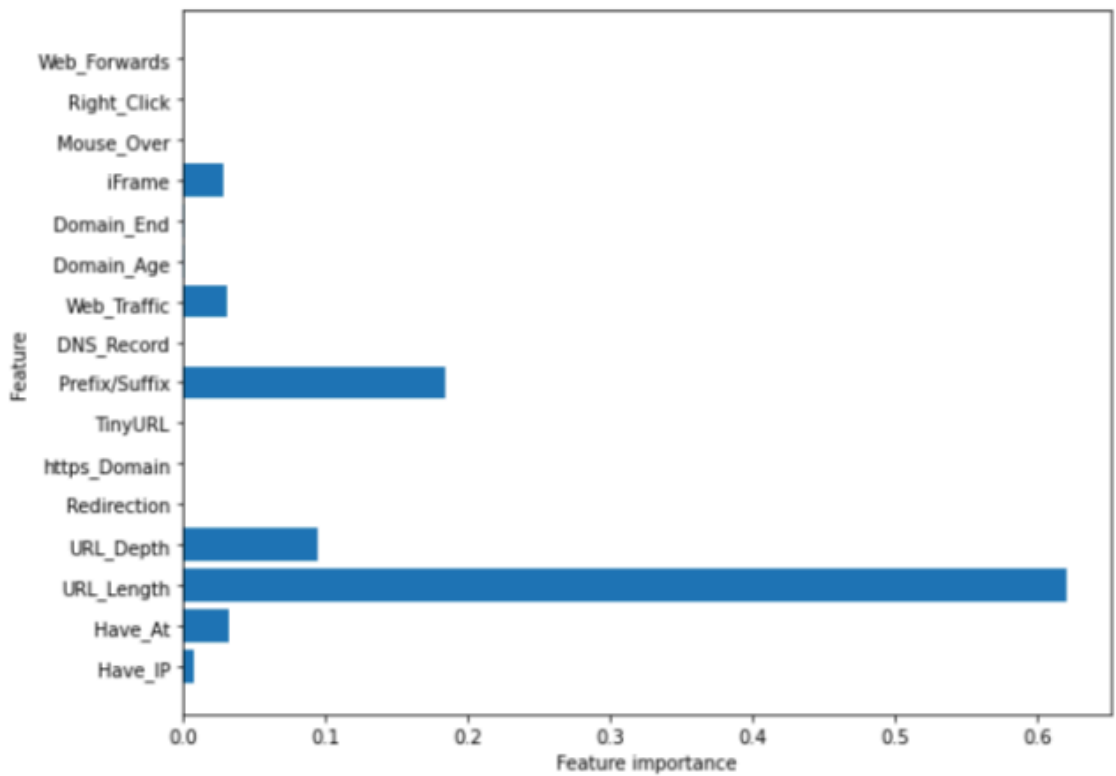

Figure 2.9: Program & Output 1.2

## 2. Random Forest Classifier

Regression and classification using random forests are two of the most popular machine learning techniques available today.In essence, a random forest is a group of decision trees, with tiny variations among the trees. Random forests work on the assumption that while individual trees may perform rather well in predicting, they will almost certainly overfit some portion of the data.

By averaging the outcomes of multiple well-functioning trees that overfit in diverse ways, we can lower the degree of overfitting. Selecting the number of trees to construct (the n_estimators option of RandomForestRegressor or RandomForestClassifier) is necessary before you can begin building a random forest model. They frequently function effectively without requiring much parameter adjustment, are incredibly strong, and don't require scaling of the data.

```python
# Random Forest model
from sklearn.ensemble import RandomForestClassifier

# instantiate the model
forest = RandomForestClassifier(max_depth=5)

# fit the model
forest.fit(X_train, y_train)
```

```
        RandomForestClassifier
RandomForestClassifier(max_depth=5)
```

```python
#predicting the target value from the model for the samples
y_test_forest = forest.predict(X_test)
y_train_forest = forest.predict(X_train)
```

Figure 2.10: Program & Output 1.3

```
#computing the accuracy of the model performance
acc_train_forest = accuracy_score(y_train,y_train_forest)
acc_test_forest = accuracy_score(y_test,y_test_forest)

print("Random forest: Accuracy on training Data: {:.3f}".format(acc_train_forest))
print("Random forest: Accuracy on test Data: {:.3f}".format(acc_test_forest))
```

Random forest: Accuracy on training Data: 0.814
Random forest: Accuracy on test Data: 0.834

```
#checking the feature improtance in the model
plt.figure(figsize=(9,7))
n_features = X_train.shape[1]
plt.barh(range(n_features), forest.feature_importances_, align='center')
plt.yticks(np.arange(n_features), X_train.columns)
plt.xlabel("Feature importance")
plt.ylabel("Feature")
plt.show()
```

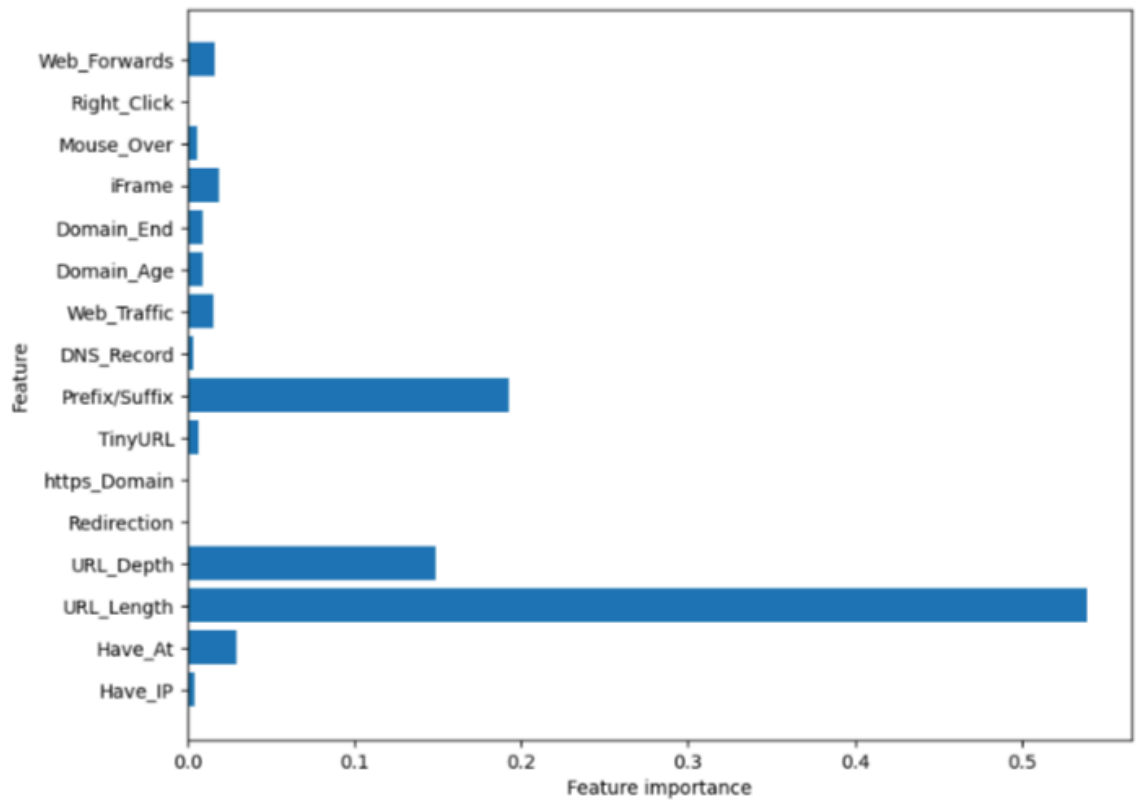

Figure 2.11: Program & Output 1.4

## 3. Multilayer Perceptrons (MLPs): Deep Learning

Multilayer perceptrons (MLPs) are sometimes referred to as neural networks, or alternatively as (vanilla) feed-forward neural networks. You can use multilayer perceptrons for regression as well as classification tasks.

MLPs are generalizations of linear models that process information via several steps in order to reach a conclusion.

```python
# Multilayer Perceptrons model
from sklearn.neural_network import MLPClassifier

# instantiate the model
mlp = MLPClassifier(alpha=0.001, hidden_layer_sizes=([100,100,100]))

# fit the model
mlp.fit(X_train, y_train)
```

```
                         MLPClassifier
MLPClassifier(alpha=0.001, hidden_layer_sizes=[100, 100, 100])
```

```python
#predicting the target value from the model for the samples
y_test_mlp = mlp.predict(X_test)
y_train_mlp = mlp.predict(X_train)
```

```python
#computing the accuracy of the model performance
acc_train_mlp = accuracy_score(y_train,y_train_mlp)
acc_test_mlp = accuracy_score(y_test,y_test_mlp)

print("Multilayer Perceptrons: Accuracy on training Data: {:.3f}".format(acc_train_mlp))
print("Multilayer Perceptrons: Accuracy on test Data: {:.3f}".format(acc_test_mlp))
```

```
Multilayer Perceptrons: Accuracy on training Data: 0.865
Multilayer Perceptrons: Accuracy on test Data: 0.865
```

Figure 2.12: Program & Output 1.5

### 4. XGBoost Classifier

One of the most widely used machine learning algorithms nowadays is XGBoost. The acronym for eXtreme Gradient Boosting is XGBoost. Whichever kind of prediction task—classification or regression—is being performed. A gradient boosted decision tree solution created for speed and efficiency is called XGBoost.

```python
#XGBoost Classification model
from xgboost import XGBClassifier

# instantiate the model
xgb = XGBClassifier(learning_rate=0.4,max_depth=7)
#fit the model
xgb.fit(X_train, y_train)
```

```
                        XGBClassifier
XGBClassifier(base_score=None, booster=None, callbacks=None,
              colsample_bylevel=None, colsample_bynode=None,
              colsample_bytree=None, device=None, early_stopping_rounds=None,
              enable_categorical=False, eval_metric=None, feature_types=None,
              gamma=None, grow_policy=None, importance_type=None,
              interaction_constraints=None, learning_rate=0.4, max_bin=None,
              max_cat_threshold=None, max_cat_to_onehot=None,
              max_delta_step=None, max_depth=7, max_leaves=None,
              min_child_weight=None, missing=nan, monotone_constraints=None,
              multi_strategy=None, n_estimators=None, n_jobs=None,
              num_parallel_tree=None, random_state=None, ...)
```

```python
#predicting the target value from the model for the samples
y_test_xgb = xgb.predict(X_test)
y_train_xgb = xgb.predict(X_train)
```

Figure 2.13: Program & Output 1.6

An auto encoder is a neural network that has the same number of input neurons as it does outputs. The hidden layers of the neural network will have fewer neurons than the input/output neurons. Because there are fewer neurons, the auto-encoder must learn to encode the input to the fewer hidden neurons. The predictors (x) and output (y) are exactly the same in an auto encoder.

```
#computing the accuracy of the model performance
acc_train_xgb = accuracy_score(y_train,y_train_xgb)
acc_test_xgb = accuracy_score(y_test,y_test_xgb)

print("XGBoost: Accuracy on training Data: {:..3f}".format(acc_train_xgb))
print("XGBoost : Accuracy on test Data: {:..3f}".format(acc_test_xgb))
```

```
XGBoost: Accuracy on training Data: 0.867
XGBoost : Accuracy on test Data: 0.864
```

Figure 2.14: Program & Output 1.7

## 5. Autoencoder Neural Network

```python
#importing required packages
import keras
from keras.layers import Input, Dense
from keras import regularizers
import tensorflow as tf
from keras.models import Model
from sklearn import metrics
```

Using TensorFlow backend.

```python
#building autoencoder model

input_dim = X_train.shape[1]
encoding_dim = input_dim

input_layer = Input(shape=(input_dim, ))
encoder = Dense(encoding_dim, activation="relu",
                activity_regularizer=regularizers.l1(10e-4))(input_layer)
encoder = Dense(int(encoding_dim), activation="relu")(encoder)

encoder = Dense(int(encoding_dim-2), activation="relu")(encoder)
code = Dense(int(encoding_dim-4), activation='relu')(encoder)
decoder = Dense(int(encoding_dim-2), activation='relu')(code)

decoder = Dense(int(encoding_dim), activation='relu')(encoder)
decoder = Dense(input_dim, activation='relu')(decoder)
autoencoder = Model(inputs=input_layer, outputs=decoder)
autoencoder.summary()
```

```
Model: "model_1"
_________________________________________________________________
Layer (type)                 Output Shape              Param #
=================================================================
input_1 (InputLayer)         (None, 16)                0
_________________________________________________________________
dense_1 (Dense)              (None, 16)                272
_________________________________________________________________
dense_2 (Dense)              (None, 16)                272
_________________________________________________________________
dense_3 (Dense)              (None, 14)                238
_________________________________________________________________
dense_6 (Dense)              (None, 16)                240
_________________________________________________________________
dense_7 (Dense)              (None, 16)                272
=================================================================
Total params: 1,294
Trainable params: 1,294
Non-trainable params: 0
_________________________________________________________________
```

Figure 2.15: Program & Output 1.8

```
#compiling the model
autoencoder.compile(optimizer='adam',
                    loss='binary_crossentropy',
                    metrics=['accuracy'])

#Training the model
history = autoencoder.fit(X_train, X_train, epochs=10, batch_size=64, shuffle=True, validation_split=0.2)

Train on 6400 samples, validate on 1600 samples
Epoch 1/10
6400/6400 [==============================] - 0s 51us/step - loss: 1.3997 - accuracy: 0.7132 - val_loss: -0.3941 - val_accuracy: 0.7890
Epoch 2/10
6400/6400 [==============================] - 0s 24us/step - loss: -0.4269 - accuracy: 0.7821 - val_loss: -0.5190 - val_accuracy: 0.7812
Epoch 3/10
6400/6400 [==============================] - 0s 24us/step - loss: -1.0514 - accuracy: 0.7908 - val_loss: -1.3147 - val_accuracy: 0.8149
Epoch 4/10
6400/6400 [==============================] - 0s 24us/step - loss: -1.3118 - accuracy: 0.8200 - val_loss: -1.3532 - val_accuracy: 0.8128
Epoch 5/10
6400/6400 [==============================] - 0s 25us/step - loss: -1.3789 - accuracy: 0.8168 - val_loss: -1.4710 - val_accuracy: 0.8190
Epoch 6/10
6400/6400 [==============================] - 0s 25us/step - loss: -1.4435 - accuracy: 0.8187 - val_loss: -1.5160 - val_accuracy: 0.8204
Epoch 7/10
6400/6400 [==============================] - 0s 25us/step - loss: -1.4951 - accuracy: 0.8215 - val_loss: -1.5601 - val_accuracy: 0.8240
Epoch 8/10
6400/6400 [==============================] - 0s 23us/step - loss: -1.5208 - accuracy: 0.8192 - val_loss: -1.5912 - val_accuracy: 0.8236
Epoch 9/10
6400/6400 [==============================] - 0s 25us/step - loss: -1.5044 - accuracy: 0.8140 - val_loss: -1.5868 - val_accuracy: 0.8191
Epoch 10/10
6400/6400 [==============================] - 0s 25us/step - loss: -1.5554 - accuracy: 0.8214 - val_loss: -1.6153 - val_accuracy: 0.8205
```

```
acc_train_auto = autoencoder.evaluate(X_train, X_train)[1]
acc_test_auto = autoencoder.evaluate(X_test, X_test)[1]

print('\nAutoencoder: Accuracy on training Data: {:.3f}' .format(acc_train_auto))
print('Autoencoder: Accuracy on test Data: {:.3f}' .format(acc_test_auto))

8000/8000 [==============================] - 0s 18us/step
2000/2000 [==============================] - 0s 20us/step

Autoencoder: Accuracy on training Data: 0.819
Autoencoder: Accuracy on test Data: 0.818
```

Figure 2.16: Program & Output 1.9

### 6. Support Vector Machines

Support-vector machines, also known as support-vector networks in machine learning, are supervised learning models that examine data used for regression and classification analysis along with corresponding learning methods. An SVM training procedure creates a model that assigns new examples to one or the other category given a collection of training examples that are all designated as belonging to one or the other of two categories. This makes the model a non-probabilistic binary linear classifier.

```python
#Support vector machine model
from sklearn.svm import SVC

# instantiate the model
svm = SVC(kernel='linear', C=1.0, random_state=12)
#fit the model
svm.fit(X_train, y_train)
```

```
SVC(C=1.0, break_ties=False, cache_size=200, class_weight=None, coef0=0.0,
    decision_function_shape='ovr', degree=3, gamma='scale', kernel='linear',
    max_iter=-1, probability=False, random_state=12, shrinking=True, tol=0.001,
    verbose=False)
```

```python
#predicting the target value from the model for the samples
y_test_svm = svm.predict(X_test)
y_train_svm = svm.predict(X_train)
```

```python
#Support vector machine model
from sklearn.svm import SVC

# instantiate the model
svm = SVC(kernel='linear', C=1.0, random_state=12)
#fit the model
svm.fit(X_train, y_train)
```

```
SVC(C=1.0, break_ties=False, cache_size=200, class_weight=None, coef0=0.0,
    decision_function_shape='ovr', degree=3, gamma='scale', kernel='linear',
    max_iter=-1, probability=False, random_state=12, shrinking=True, tol=0.001,
    verbose=False)
```

```python
#predicting the target value from the model for the samples
y_test_svm = svm.predict(X_test)
y_train_svm = svm.predict(X_train)
```

```python
#computing the accuracy of the model performance
acc_train_svm = accuracy_score(y_train,y_train_svm)
acc_test_svm = accuracy_score(y_test,y_test_svm)

print("SVM: Accuracy on training Data: {:.3f}".format(acc_train_svm))
print("SVM : Accuracy on test Data: {:.3f}".format(acc_test_svm))
```

```
SVM: Accuracy on training Data: 0.798
SVM : Accuracy on test Data: 0.818
```

Figure 2.17: Program & Output 2.0

**Comparison of Models**

| | ML Model | Train Accuracy | Test Accuracy |
|---|---|---|---|
| 0 | Decision Tree | 0.810 | 0.826 |
| 1 | Random Forest | 0.814 | 0.834 |
| 2 | Multilayer Perceptrons | 0.858 | 0.863 |
| 3 | XGBoost | 0.866 | 0.864 |
| 4 | AutoEncoder | 0.819 | 0.818 |
| 5 | SVM | 0.798 | 0.818 |

Figure 2.18: Comparison of Models

For the above comparison, it is clear that the XGBoost Classifier works well with this dataset.

# CHAPTER 3
# TECHNOLOGY ANALYSIS

**3.1 UML Diagram**

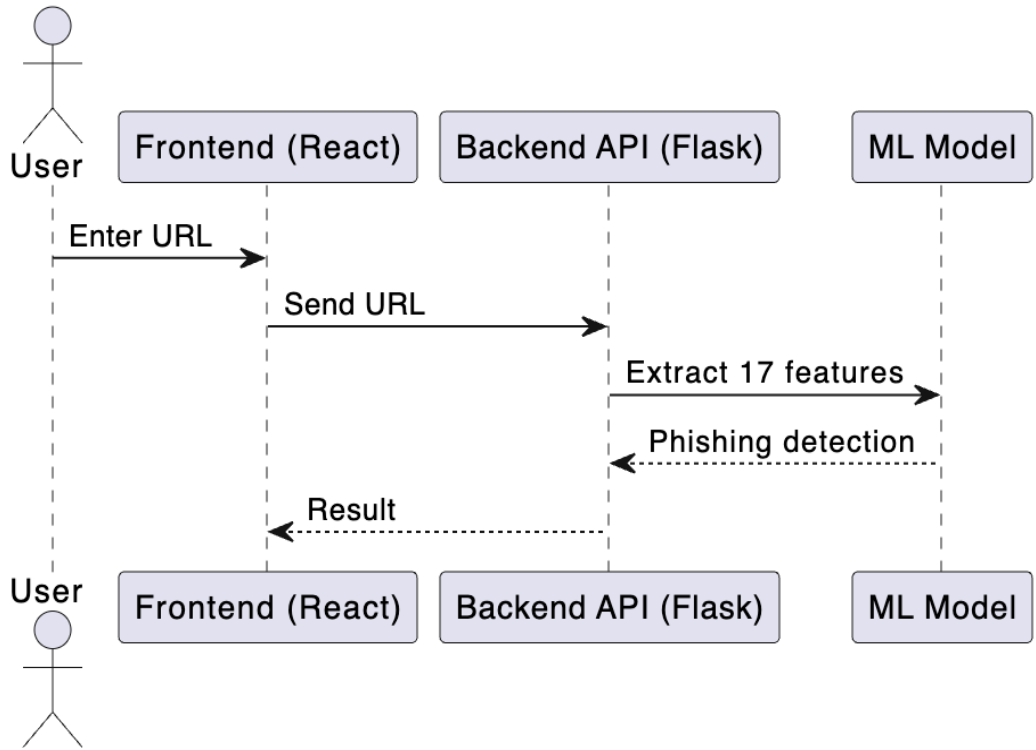

Figure 3.1: UML Diagram

**3.2 Tech Stack Analysis**

Tech stack analysis allows teams to assess the compatibility and effectiveness of the tools and technologies being used. This analysis helps identify potential bottlenecks, inefficiencies, or areas where newer technologies could offer better solutions. By optimizing the tech stack, teams can streamline development processes, improve product quality, and enhance overall project success.

**3.2.1 ML Model and API Services:**

Python and Flask have been used for machine learning models and Application Programming Interfaces.It offers a robust and flexible framework for developing and deploying machine learning applications. Python's extensive libraries, such as TensorFlow, Scikit-learn, and PyTorch, provide powerful tools for building and training machine learning models. Flask, with its lightweight and minimalist design, offers a straightforward way to create RESTful APIs, enabling seamless communication between the frontend and backend systems. Flask's simplicity allows developers to quickly get started with web development while providing flexibility for customization and extension through its modular design. This combination allows for the integration of machine learning functionalities into web applications, enabling tasks such as prediction, classification, and data analysis to be performed in real- time.

**3.2.2 Frontend Framework :**

React is a JavaScript library developed by Facebook for building user interfaces, particularly for single-page applications (SPAs) and dynamic web applications. It's known for its component-based architecture, which promotes reusability, modularity, and maintainability in frontend development. Developers can create UI components that encapsulate their own state and behavior, making it easier to manage complex UIs. These components can be composed together to build larger UIs, providing a flexible and scalable approach to frontend development. React's virtual DOM (Document Object Model) efficiently updates and renders UI components, leading to better performance and a smoother user experience. It also encourages the use of a unidirectional data flow, where data flows from parent components to child components through props, and updates are handled using state. This paradigm simplifies data management and ensures predictable behavior in the application.

**3.2.3 R&D and Comparison of Models :**

Google Colab, a cloud-based Jupyter notebook environment, is a valuable tool for research and development (R&D) in machine learning and data science, particularly for comparing models. Google Colab provides free access to GPU and TPU resources, which are crucial for training complex machine learning models efficiently. This accessibility eliminates the need for expensive hardware investments, making it ideal for R&D purposes where experimentation is frequent. It allows multiple users to work on the same notebook simultaneously, enabling collaboration among researchers and data scientists. This collaborative environment fosters knowledge sharing, facilitates peer reviews, and accelerates the development process. Colab supports Jupyter notebooks, which provide an interactive and exploratory environment for writing code, documenting experiments, and visualizing results. Researchers can easily experiment with different algorithms, hyperparameters, and datasets within the same notebook, enabling rapid iteration and comparison of models.

**3.2.4 Backend Framework :**

Flask is a lightweight and versatile web framework for Python, which is commonly used as a backend framework for building web applications and APIs. It follows a micro-framework approach, meaning it provides only the important features needed for web development, while allowing developers to add additional functionality as needed. This flexibility enables developers to tailor the framework to the specific requirements of their project, without being constrained by unnecessary features. It is built around the concept of modular components, known as extensions. These extensions provide additional functionality such as database integration, authentication, and RESTful API support, allowing developers to easily add features to their Flask applications without rebuilding the wheel. Developers can use Flask to create APIs that expose data and functionality to client applications, enabling seamless communication between frontend and backend systems.

**3.2.5 Chrome Extension :**

Vanilla JavaScript for Chrome extension development offers a lightweight and efficient approach without the need for additional libraries or frameworks. Its compatibility across various browsers ensures a broader reach and reduces the likelihood of compatibility issues. With a focus on performance, vanilla JavaScript enables fast and responsive interactions with the browser, crucial for providing a seamless user experience. Its simplicity makes it ideal for developers of all levels, facilitating a deeper understanding of browser APIs and DOM manipulation. The flexibility of vanilla JavaScript allows developers to implement custom solutions tailored to the specific functionality and requirements of the Chrome extension, promoting creativity and innovation in extension development.

# CHAPTER 4
# ECONOMIC ANALYSIS

We have built our applications and platforms using free and secure technologies, including APIs, datasets, and dependencies. As these are also free software, they only require support and a willingness to adapt to any changes that may arise. So, we guarantee that there will be no costs that are associated with using our application, and all requirements will be completely free.

- Our goal is to offer solutions that are affordable, user-friendly, and equipped with the necessary features to tackle common challenges.
- As for the various development stacks used, they are freely available and require an internet connection.
- Google Colab is free to use.
- AWS subscription is free up to a certain time and the basic EC2 instance resources are enough to train the model for free.
- The use of freely available technologies enables businesses to scale their operations without incurring proportional increases in software-related expenses.
- Free software often receives updates and improvements from the community, ensuring that applications and platforms remain current and competitive without additional financial investments.
- It has a long term affordability as there is no cost associated with using the application.
- Investing in free, open-source technologies increases long-term sustainability by reducing dependency on software vendors.

# CHAPTER 5

# RESULT AND DISCUSSION

## 5.1. Website Usage Instructions

- The dashboard of the application allows the users to input the url that they need to check. Then they are given the options to either check or report the website.

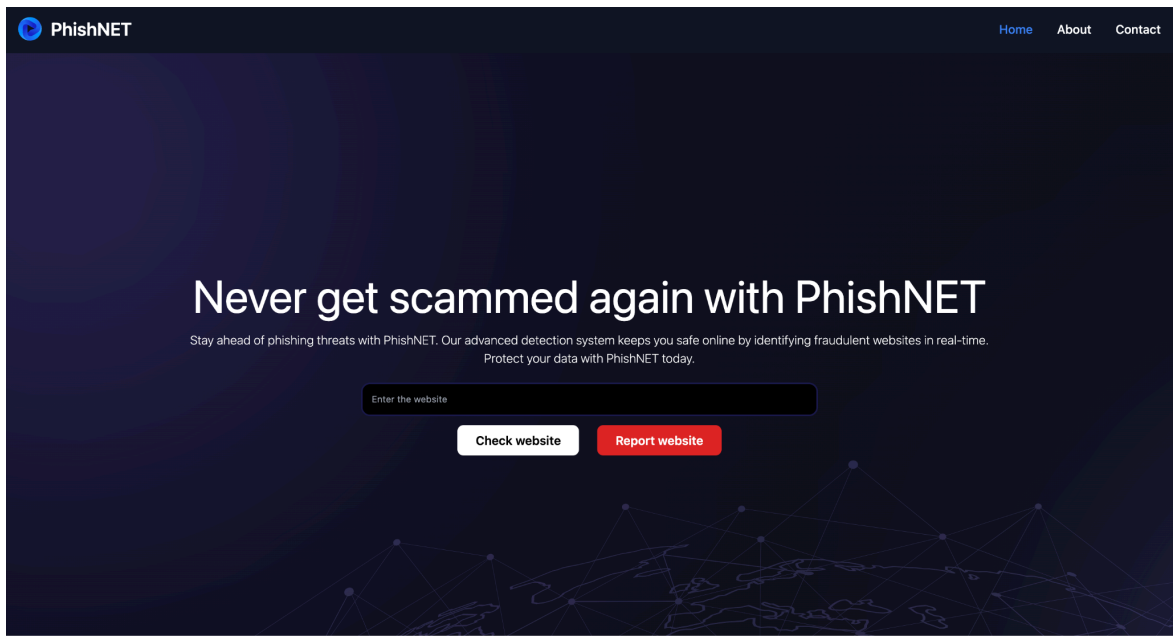

Figure 5.1: Landing Page

- Once Reaching the Landing Page, the user views a text box where they can pase any URL they want to test this on.

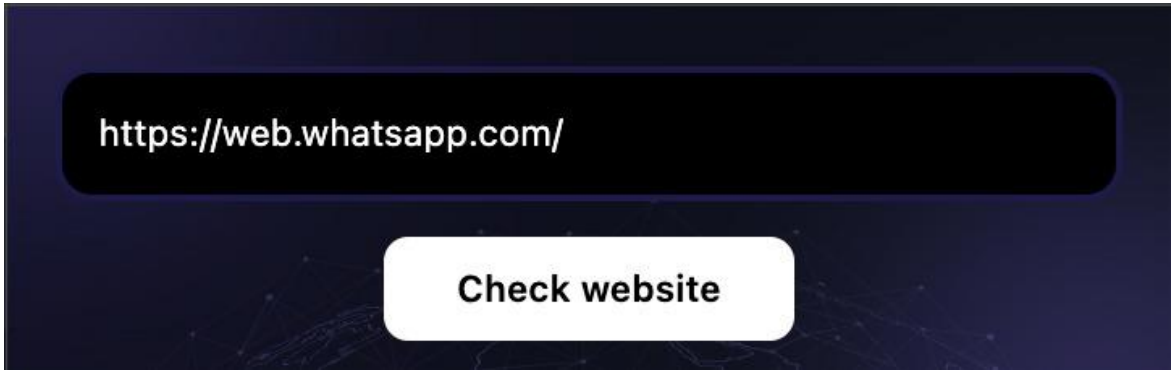

Figure 5.2: URL Search

- When the user enters a URL which is predicted to be a phishing URL and clicks on "Check Website", a red pop up warns the user that the URL is a phishing URL.

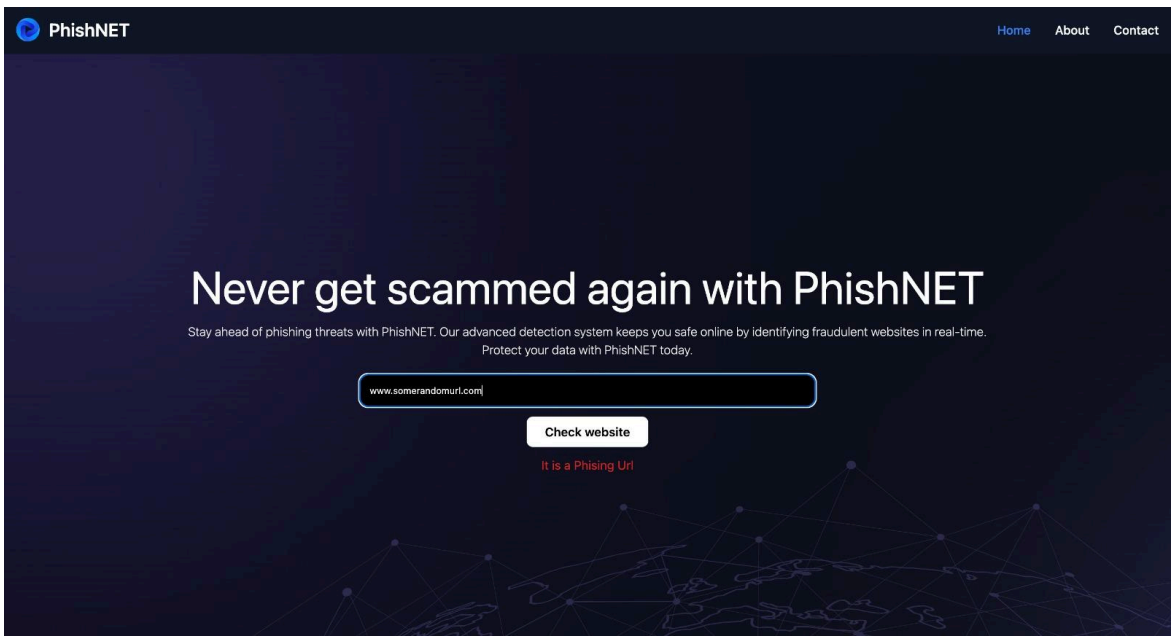

Figure 5.3: Phishing URL

- When the user enters a URL which is predicted to be a legitimate URL and clicks on "Check Website", a red pop up notifies the user that the URL is a legitimate URL.

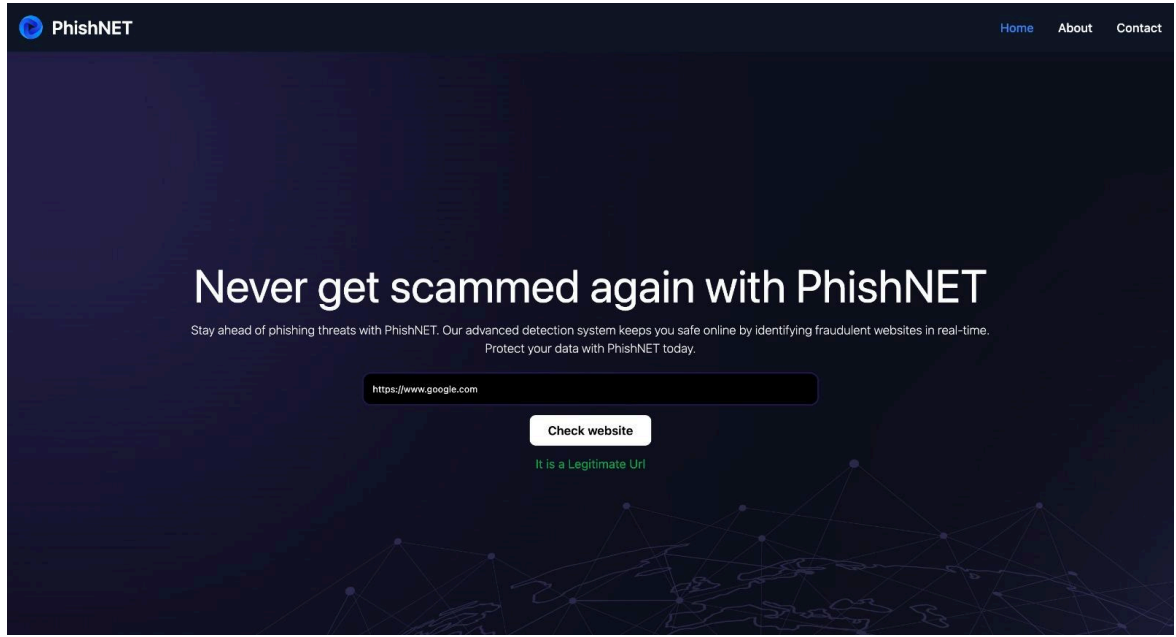

Figure 5.4: Legitimate URL

## 5.2 Risk Analysis

Designing or developing anything can't be risk-free. Risks can occur, but with a proper risk mitigation, monitoring, and management plan, we can make our project resistant to any adverse consequences, should any risk occur. Hereby, various risks are highlighted that are involved in our project:

1. Data Privacy and Security Risks: Collecting and preprocessing a comprehensive dataset of URLs for training the machine learning model may expose sensitive information, potentially leading to privacy breaches or data leaks. Adequate measures must be implemented to anonymize or encrypt data to mitigate these risks.
2. Model Performance Risks: While multiple machine learning algorithms are evaluated for their performance, there is a risk that the selected algorithms may not accurately detect phishing websites in all scenarios. Variations in phishing tactics and the evolving nature of cyber threats pose challenges in achieving consistent model performance

3. False Positive and False Negative Risks: Despite meticulous training to optimize performance metrics, there is a risk of false positives (legitimate websites incorrectly classified as phishing) and false negatives (phishing websites incorrectly classified as legitimate). These errors can undermine user trust in the application and lead to potential security vulnerabilities.
4. Dependency Risks: PhisNet relies on external dependencies, including libraries, frameworks, and cloud services, which may introduce risks related to version compatibility, security vulnerabilities, or service disruptions. Regular updates and risk assessments of dependencies are necessary to mitigate these risks and ensure the application's stability and security.
5. Backend Infrastructure Risks: The robustness and scalability of the backend infrastructure, deployed on Google Colab and AWS EC2, are critical for ensuring the application remains accessible and functional under varying loads of user requests. Downtime, network outages, or resource limitations could disrupt service availability and impact user experience.

# CHAPTER 6
# CONCLUSION

PhisNet represents a significant advancement in the ongoing battle against phishing attacks, providing users with powerful tools to detect and mitigate potential threats. Through the development of a web application and a Chrome extension, PhisNet offers accessible and effective solutions for identifying phishing URLs in real-time.

As phishing attacks continue to evolve in sophistication, it is essential to deploy proactive measures to safeguard against these threats. PhisNet's utilization of advanced machine learning algorithms enables users to stay one step ahead of cybercriminals by accurately identifying and avoiding phishing attempts.

It is crucial to emphasize that PhisNet does not seek to replace the expertise of cybersecurity professionals but rather to complement their efforts. While PhisNet provides users with preliminary assessments of URL legitimacy, users should always exercise caution and seek professional guidance when encountering suspicious online content.

In conclusion, PhisNet serves as a valuable asset in the ongoing fight against phishing attacks, empowering users to protect themselves and their sensitive information while navigating the digital landscape. Through continuous refinement and adaptation, PhisNet will remain at the forefront of cybersecurity innovation, ensuring users can browse the internet with confidence and security.